\documentclass[12pt,reqno]{amsart}
\usepackage{amsfonts, amsmath, amssymb, amsthm, amsbsy}

\newtheorem{theor}{Theorem}

\newtheorem{state}{Proposition}
\theoremstyle{definition}

\theoremstyle{remark}
\newtheorem{rem}{Remark}
\newtheorem{example}{Example}

\newcommand{\Id}{{\mathrm d}}

\newcommand{\BBR}{{\mathbb{R}}}
\newcommand{\BBN}{{\mathbb{N}}}

\newcommand{\cE}{\mathcal{E}}

\newcommand{\cH}{\mathcal{H}}

\newcommand{\cU}{\mathcal{U}}
\newcommand{\cV}{\mathcal{V}}

\newcommand{\vph}{\varphi}
\newcommand{\veps}{\varepsilon}

\newcommand{\tu}{\tilde{u}}
\newcommand{\tv}{\tilde{v}}

\newcommand{\by}[1]{\textit{{#1}}}
\newcommand{\jour}[1]{\textit{{#1}}}
\newcommand{\vol}[1]{\textbf{{#1}}}
\newcommand{\book}[1]{\textrm{{#1}}}

\begin{document}
   \rightline{ISPUmath-1/2006}

\title{Gardner's deformations of the Boussinesq equations}
\date{March 21, 2006}

\author[A.\,Karasu]{Atalay Karasu}

\address{Department of Physics, Middle East Technical University,
06531 Ankara, Turkey.}
\email{karasu@metu.edu.tr}

\author[A.\,V.\,Kiselev]{Arthemy\,V.\,Kiselev}

\address{
\textup{\textit{Permanent address} (A.\,V.\,K.):}
Department of Higher Mathematics,\ Ivanovo State
Power University,\ Rabfakovskaya str.\,34, Ivanovo, 153003
Russia.}

\curraddr{
Department of Physics, Middle East Technical University,
06531 Ankara, Turkey.}

\email{arthemy@newton.physics.metu.edu.tr}

\subjclass[2000]{
  35Q53, 
  37K05, 
  37K10, 
  37K35. 
\textit{PACS} 02.30.Ik. 
}

\keywords{Gardner's deformations, integrable extensions,
Boussinesq equation, Kaup\/-\/Boussinesq equation}

\begin{abstract}
Using the algebraic method of Gardner's deformations for completely
integrable systems, we construct the recurrence relations
for densities of the Hamiltonians for the Boussinesq and the
Kaup\/-\/Boussinesq equations.
By extending the Magri schemes for these equations, we obtain new
integrable systems adjoint with respect to the initial ones and describe
their Hamiltonian structures and symmetry properties.
\end{abstract}

\maketitle

\subsection*{Introduction}
In this paper we consider the most efficient way to prove the complete
integrability of evolutionary systems. Namely, we apply the method of
Gardner's deformations~\cite{Gardner}--\cite{Andrea}
to the Boussinesq and Kaup-Boussinesq equations.
By construction, the deformations consist of the (multi-)\/Hamiltonian
scaling non\/-\/invariant parametric extensions~$\cE_\veps$ of the original
systems~$\cE_0$ and the parameter\/-\/dependent
Miura transformations~$\cE_\veps\to\cE_0$.
Inverting the Miura transformations from the new systems, one
obtains the recurrence relations for the infinite sequences of
densities of the Hamiltonian functionals.
We improve the result of Kupershmidt in~\cite{KuperIrish} by showing
that the sequences of conserved densities for the Boussinesq
equation satisfy two different recurrence relations simultaneously.
We demonstrate that the extended equations must not necessarily
interpolate (as it is assumed in~\cite{MathieuN=2}) between
the equations~$\cE_0$
and the modified systems that provide the canonical factorizations
for the higher Poisson structures~\cite{Manin}.
The recurrence relations we obtain are inherited by the modified
Boussinesq~\cite{PavlovFPM} and Kaup\/--\/Boussinesq~\cite{PavlovJNlin}
equations.

Next, we separate the flows at the deformation parameters and
determine the adjoint Boussinesq equations whose Magri schemes are
coupled with the non\/-\/extended ones. We show
that the new systems are also integrable.

The examples discussed in this paper are essentially used in the
general algebraic approach to the problem of Gardner's deformations
and integrable extensions of the Magri schemes. This approach is based
on the notion of coverings over PDE~\cite{ClassSym} and their
parametric families constructed using the Fr\"olicher\/--\/Nijenhuis
bracket~\cite{JKIgonin, DeformLiou, NonAbel}. This will be the object
of a subsequent publication.

The paper is organized as follows.
In section~\ref{SecBous} we obtain two Gardner's deformations for the
Boussinesq equation using the dispersionless
case~\cite{Kiev2005} as the starting point. We specify the adjoint
Boussinesq systems and discuss their integrability.
In section~\ref{SecKB} we construct the deformation
of the Kaup\/-\/Boussinesq equation and describe the bi\/-\/Hamiltonian
adjoint system.

\section{The Boussinesq equation}\label{SecBous}
First we consider the Boussinesq equation with dispersion and
dissipation, see~\cite{KuperIrish},
\begin{equation}\label{BousDissip}
\cU_t=\cV_x+\alpha\,\cU_{xx},\quad
\cV_t=\cU_{xxx}+\cU\cU_x-\alpha\cV_{xx},\qquad \alpha\in\BBR.
\end{equation}
We note that for any~$\alpha$ system~\eqref{BousDissip} is
transformed to the equation
$\cU_{tt}=\bigl(\cU_{xxx}\cdot(1+\alpha)+\cU\cU_x\bigr)_x$,
which is scaling\/-\/equivalent to the Boussinesq equation
$u_{tt}=\bigl(u_{xxx}+uu_x\bigr)_x$ whenever~$\alpha\neq-1$.
Therefore for $\alpha\neq-1$
equation~\eqref{BousDissip} is reduced to the Boussinesq equation
\begin{equation}\label{Bous}
u_t=v_x,\qquad v_t=u_{xxx}+uu_x.
\end{equation}
In the sequel, we consider the problem of Gardner's deformation for
system~\eqref{Bous}.  

Next, we observe that a Gardner's deformation for~\eqref{Bous} (and
for the Kaup\/--\/Boussinesq equation~\eqref{KB} as well) is obtained
from the deformation of the dispersionless reduction by adding higher
order terms to the equations~$\cE_\veps$
and to the Miura transformations~$\cE_\veps\to\cE_0$.
Indeed, the zero order terms in the conserved densities, which are
obtained recursively by inverting the Miura substitutions,
originate from the relations for the dispersionless systems.
The deformation for the dispersionless Boussinesq equation has been
described in~\cite{Kiev2005}; it corresponds to the case $\alpha=-1$
in~\eqref{BousDissip}.
Using this result, we construct two deformations
for~\eqref{Bous} and get two distinct recurrence relations for both
sequences of the conserved densities.
Thus we improve the result of~\cite{KuperIrish}, where a recurrence
relation for only one sequence was obtained.
The adjoint Boussinesq equations that appear in the extended systems
are the by\/-\/products of our reasonings.

\begin{theor}
There are two deformations~$\cE_\veps^{\pm}$
for the Boussinesq equation~\eqref{Bous}.
Both of them are Hamiltonian w.r.t.\ the structure
\begin{equation}\label{FirstStr}
\left(\begin{array}{cc} 0 & D_x \\ D_x & 0\end{array}
\right)
\end{equation}
and the functionals with the densities
\[
H_{\pm}(\veps)=
   \frac{ 1}{ 6}\tu^3 + \frac{ 1}{ 2}\tv^2
   +\veps^3\cdot\Bigl(\frac{ 1}{ 6}\tv^3
    + \frac{ 1}{ 2}\tu_x^2\tv \pm \frac{ 1}{ 2}\tu_x\tv^2
    \pm \frac{ 1}{ 6}\tu_x^3\Bigr),
\]
respectively.

The deformed equation $\cE_\varepsilon^{+}$ is
\begin{eqnarray}\nonumber
\tu_t&=&\tv_x+\varepsilon^3\cdot\bigl(
   \tu_x\tu_{xx}+\tu_{xx}\tv+\tu_x\tv_x+\tv\tv_x\bigr),\\
\tv_t&=&\tu_{xxx}+\tu\tu_x - \varepsilon^3\cdot\bigl(
   \tu_{xx}^2 + \tu_x\tu_{xxx} + \tu_{xxx}\tv + 2\tu_{xx}\tv_x
   + \tu_x\tv_{xx}+\tv_x^2+\tv\tv_{xx}\bigr).\label{BousEP}
\end{eqnarray}
The Miura transformation from $\cE_\varepsilon^{+}$ to~\eqref{Bous} is
\begin{eqnarray*} 
u&=&\tu-2\varepsilon \tu_x + 2\veps^2\cdot(\tu_{xx}+\tv_x)
   + \veps^3\cdot(\tu\tv+\tu\tu_x),\\
v&=&\tv - 2\veps \tv_x + 2\veps^2\cdot(\tv_{xx}+\tu_{xxx}+\tu\tu_x)
   + \veps^3\cdot\Bigl(\frac{1}{3}\tu^3 + \tv^2 + \tu\tu_{xx}
     + \tv\tv_x + \tu_x\tv \Bigr)\\
 {}&&{}\qquad
   - 2\veps^4\cdot(\tu_x\tu_{xx}+\tu_x\tv_x+\tu_{xx}\tv+\tv\tv_x)
   + \veps^6\cdot\Bigl(\frac{1}{3}\tv^3 + \frac{1}{3}\tu_x^3
     + \tu_x^2\tv + \tu_x\tv^2\Bigr).
\end{eqnarray*}

The deformed equation $\cE_\varepsilon^{-}$ is
\begin{eqnarray}\nonumber
\tu_t&=&\tv_x+\varepsilon^3\cdot\bigl(
   \tu_x\tu_{xx}-\tu_{xx}\tv-\tu_x\tv_x+\tv\tv_x\bigr),\\
\tv_t&=&\tu_{xxx}+\tu\tu_x
  + \varepsilon^3\cdot\bigl(
   \tu_{xx}^2 + \tu_x\tu_{xxx}-\tu_{xxx}\tv - 2\tu_{xx}\tv_x
   -\tu_x\tv_{xx}+\tv_x^2+\tv\tv_{xx}\bigr).\label{BousEM}
\end{eqnarray}
The Miura transformation from $\cE_\varepsilon^{-}$ to~\eqref{Bous}
is given through
\begin{eqnarray*}   
 u&=&\tu+2\varepsilon \tu_x + 2\veps^2\cdot(\tu_{xx}-\tv_x)
   + \veps^3\cdot(\tu\tv-\tu\tu_x),\\
 v&=&\tv + 2\veps \tv_x + 2\veps^2\cdot(\tv_{xx}-\tu_{xxx}-\tu\tu_x)
   + \veps^3\cdot\Bigl(\frac{1}{3}\tu^3 + \tv^2 + \tu\tu_{xx}
     - \tv\tv_x - \tu_x\tv \Bigr)\\
 {}&&{}\qquad
   + 2\veps^4\cdot(\tu_x\tu_{xx}-\tu_x\tv_x-\tu_{xx}\tv+\tv\tv_x)
   + \veps^6\cdot\Bigl(\frac{1}{3}\tv^3 - \frac{1}{3}\tu_x^3
     + \tu_x^2\tv - \tu_x\tv^2\Bigr).
\end{eqnarray*}
\end{theor}

Extended equations~\eqref{BousEP} and~\eqref{BousEM}
consist of the original Boussinesq flows and the adjoint
flows at~$\veps^3$, which will be further discussed in more detail.
The Poisson structure~\eqref{FirstStr}
for the extended equations~$\cE_\veps^{\pm}$
together with the Miura transformations
$\cE_\veps^{\pm}\to\cE_0$ induce~\cite{Manin} the Poisson structures
$\hat{A}_1\mp\veps^3\hat{A}_2$ for Boussinesq's equation~\eqref{Bous},
here $\hat{A}_1$ is given through~\eqref{FirstStr} and
\[
\hat{A}_2=\left(\begin{array}{cc}
8D_x^3 + uD_x+D_x\circ u & 3vD_x+v_x \\
3vD_x+2v_x & \begin{array}{l} D_x^5 +5\cdot(uD_x^3+D_x^3\circ u)-{}\\
  {}\quad -3\cdot(u_{xx}D_x+D_x\circ u_{xx}) + uD_x\circ u
\end{array}
\end{array}\right)
\]
are its first and second structures, respectively
(see~\cite{PavlovFPM} and references therein).
The Hamiltonians for the extensions~$\cE_\veps^{\pm}$ are inherited from
the original functionals, which are described in
Proposition~\ref{BousDenstState} below, by using the Miura substitutions.

\begin{state}\label{BousDenstState}
The densities of Hamiltonian functionals for the Boussinesq equation
can be obtained using two different recurrence relations, which are
\begin{eqnarray*}
 \lefteqn{\begin{array}{lll}
\tu_0=u,  & \tv_0=v, & \tu_1=\pm 2u_x,\\
\tv_1=\pm 2v_x, & \tu_2= 2u_{xx} \mp 2v_x, &
\tv_2= 2v_{xx} \mp 2u_{xxx} \mp 2uu_x,
   \end{array}}
 \phantom{\tu_0\,\,}&&{}\\
\tu_n&=&\pm 2D_x(\tu_{n-1}) - 2D_x^2(\tu_{n-2}) \mp 2D_x(\tv_{n-2})
   +\sum_{k+\ell=n-3}\Bigl[-\tu_k\tv_\ell \mp \tu_kD_x(\tu_\ell)\Bigr],\\
   {}&&\qquad{}\qquad n\geq3,\\
\tv_3&=&\pm 2D_x(\tv_2) \mp 2D_x^3(\tu_1) - 2D_x^2(\tv_1)
   \mp \left[uD_x(\tu_1) + \tu_1u_x\right] \\
  {}&&\qquad{}
   - \frac{1}{3}u^3 - v^2 - uu_{xx} \mp uv_x \mp u_xv,\\
\tv_n&=&\pm 2D_x(\tv_{n-1}) \mp 2D_x^3(\tu_{n-2}) - 2D_x^2(\tv_{n-2}) \\
  {}&&\qquad{}
   \mp \sum_{k+\ell=n-2}2\tu_kD_x(\tu_\ell)
   - \sum_{k+\ell+m=n-3}\frac{1}{3}\tu_k\tu_\ell \tu_m\\
  {}&&\qquad{}
   + \sum_{k+\ell=n-3}\Bigl[- \tv_k\tv_\ell - \tu_kD_x^2(\tu_\ell)
      \mp \tu_kD_x(\tv_\ell) \mp D_x(\tu_k)\tv_\ell\Bigr] \\
  {}&&\qquad{}
   + \sum_{k+\ell=n-4}2\cdot\Bigl[\pm D_x(\tu_k)D_x^2(\tu_\ell)
      + D_x(\tu_k)D_x(\tv_\ell) + D_x^2(\tu_k)\tv_\ell
      \pm \tv_kD_x(\tv_\ell)\Bigr],\\
  {}&&\qquad{}\qquad n=4,5,
\end{eqnarray*}
\begin{eqnarray*}
\tv_n&=&\pm 2D_x(\tv_{n-1}) \mp 2D_x^3(\tu_{n-2}) - 2D_x^2(\tv_{n-2}) \\
  {}&&\qquad{}
   \mp \sum_{k+\ell=n-2}2\tu_kD_x(\tu_\ell)
   - \sum_{k+\ell+m=n-3}\frac{1}{3}\tu_k\tu_\ell \tu_m\\
  {}&&\qquad{}
   + \sum_{k+\ell=n-3}\Bigl[- \tv_k\tv_\ell - \tu_kD_x^2(\tu_\ell)
      \mp \tu_kD_x(\tv_\ell) \mp D_x(\tu_k)\tv_\ell\Bigr] \\
  {}&&\qquad{}
   + \sum_{k+\ell=n-4}2\cdot\Bigl[\pm D_x(\tu_k)D_x^2(\tu_\ell)
      + D_x(\tu_k)D_x(\tv_\ell) + D_x^2(\tu_k)\tv_\ell
      \pm \tv_kD_x(\tv_\ell)\Bigr] \\
  {}&&\qquad{}
   + \sum_{k+\ell+m=n-6}\Bigl[-\frac{1}{3}\tv_k\tv_l\tv_m
      \mp \frac{1}{3}D_x(\tu_k)D_x(\tu_\ell)D_x(\tu_m) \\
  {}&&\qquad{}\qquad
      - D_x(\tu_k)D_x(\tu_\ell)\tv_m + D_x(\tu_k)\tv_\ell \tv_m\Bigr],
  \qquad n\geq6.
\end{eqnarray*}
The ambiguity of signs of the differential terms does not affect the
nontrivial conserved densities $\tu_{3k}$ and $\tv_{3k}$.\ %
The densities with subscripts $3k+1$, $3k+2$
are trivial for all~$k\geq0$.
\end{state}

The recurrence relations listed in Proposition~\ref{BousDenstState}
are also valid for the Hamiltonians of the modified Boussinesq
equation~\cite{PavlovFPM} which are obtained using the Miura substitutions
to~\eqref{Bous}. The modified fields themselves do not start
these sequences of conserved densities and thus cannot be regarded
as their negative terms $\tu_{-3}$,~$\tv_{-3}$.

Further, consider the flow at $\veps^3$ in the r.h.s.\ of
equation~\eqref{BousEM};
the second case related to~\eqref{BousEP} is analogous. We thus obtain
the system
\begin{eqnarray}\nonumber
u_\tau&=& vv_x + u_xu_{xx}-u_{xx}v - u_xv_x,\\
v_\tau&=& vv_{xx} + v_x^2 - u_xv_{xx} - 2u_{xx}v_x
   - u_{xxx}v + u_xu_{xxx} + u_{xx}^2.\label{EDpBous}
\end{eqnarray}
It is Hamiltonian w.r.t.\ the
structure~\eqref{FirstStr} and the functional with density
\[
\bar{H} = \frac{ 1}{ 6}v^3
    + \frac{ 1}{ 2}u_x^2v
    - \frac{ 1}{ 2}u_xv^2
    - \frac{ 1}{ 6}u_x^3.
\]
Note that
system~\eqref{EDpBous} is homogeneous w.r.t.\ the set of weights
$|u|=2$, $|v|=3$, $|\tau|=-5$, $|x|\equiv-1$.
We claim that equation~\eqref{EDpBous} admits
two infinite sequences of Hamiltonian symmetries
${\left({u_s}, {v_s}\right)}=\vph_{[s]}$
for all weights $|s|=-(6k+3\pm2)$, $k\in\BBN_{\geq0}$.
These flows reproduce (with proper modifications)
the scheme for the symmetries of the original Boussinesq hierarchy.
Namely, they are arranged according to the diagram
\[
\vph_{[1]}\to\vph_{[7]}\to\vph_{[13]}\to\cdots,\qquad
\vph_{[5]}\to\vph_{[11]}\to\cdots.
\]
The flow $\vph_{[5]}$ is the right\/-\/hand side
of the adjoint Boussinesq equation~\eqref{EDpBous} itself.
The two sequences of Hamiltonians for the higher symmetries
of~\eqref{EDpBous} are obtained from the functionals for
system~\eqref{BousEM} by separating the coefficients of the
highest powers of~$\veps$.

\begin{example}
The components $\vph_{[7]}^{1,2}$ of the symmetry $\vph_{[7]}$
are 
\begin{eqnarray*}
\vph_{[7]}^1&=& vv_{xxx} -  u_xv_{xxx}
   + 2v_xv_{xx} - 2u_{xx}v_{xx}
   - 2u_{xxx}v_x -  u_{4x}v +  u_xu_{4x}
   + 2u_{xx}u_{xxx},\\
\vph_{[7]}^2&=& vv_{4x} -  u_xv_{4x}
   + 3v_xv_{xxx} - 3u_{xx}v_{xxx}
   + 2v_{xx}^2 - 4u_{xxx}v_{xx}
   - 3u_{4x}v_x -  u_{5x}v \\
 {}&&\qquad{} +  u_xu_{5x} + 3u_{xx}u_{4x} + 2u_{xxx}^2.
\end{eqnarray*}
The symmetry $\vph_{[11]}$ that succeeds the equation in the hierarchy
has the components 
\begin{eqnarray*}
\vph_{[11]}^1&=& vv_{7x} - u_xv_{7x}
   + 4v_xv_{6x} - 4u_{xx}v_{6x} + 10v_{xx}v_{5x} - 10u_{xxx}v_{5x}
   + 16v_{xxx}v_{4x}\\
 {}&&\qquad{}
   - 16u_{4x}v_{4x} - 16u_{5x}v_{xxx} - 10u_{6x}v_{2x}
   - 4u_{7x}v_x - u_{8x}v + u_xu_{8x}\\
 {}&&\qquad{}
   + 4u_{xx}u_{7x} + 10u_{xxx}u_{6x} + 16u_{4x}u_{5x},\\
\vph_{[11]}^2&=& vv_{8x} - u_xv_{8x}
   + 5v_xv_{7x} - 5u_{xx}v_{7x} + 14v_{xx}v_{6x} - 14u_{xxx}v_{6x}
   + 26v_{xxx}v_{5x} \\
 {}&&\qquad{}
   - 26u_{4x}v_{5x} + 16v_{4x}^2 - 32u_{5x}v_{4x}
   - 26u_{6x}v_{xxx} - 14u_{7x}v_{xx} - 5u_{8x}v_x \\
 {}&&\qquad{}
   - u_{9x}v + u_xu_{9x}
   + 5u_{xx}u_{8x} + 14u_{3x}u_{7x} + 26u_{4x}u_{6x} + 16u_{5x}^2.
\end{eqnarray*}
\end{example}

\section{The Kaup\/--\/Boussinesq equation}\label{SecKB}
Now we construct the Gardner deformation for the
Kaup\/--\/Boussinesq equation
\begin{equation}\label{KB}
u_t=uu_x+v_x,\qquad v_t=(uv)_x+ u_{xxx}.
\end{equation}

\begin{theor}\label{KBPTh}
The integrable extension~$\cE_\veps$ of the Kaup\/--\/Boussinesq
equation~\eqref{KB} is the system
\begin{eqnarray}\nonumber
\tu_t&=&\tu\tu_x+\tv_x
   + \veps\cdot\bigl(\tu\tu_{xx}+\tu_x^2+{(\tu\tv)}_x\bigr),\\
\tv_t&=&{(\tu\tv)}_x + \tu_{xxx}
   - \veps\cdot\bigl(2\tu_x\tu_{xx}+\tu\tu_{xxx}+\tu_x\tv_x+\tu\tv_{xx}
      - \tv\tv_x\bigr).\label{KBe}
\end{eqnarray}
System~\eqref{KBe} is Hamiltonian w.r.t.\ the structure~\eqref{FirstStr}
and the functional
\[
\cH(\veps)=\int\Bigl(
   \frac{ 1}{ 2}\tu^2\tv + \frac{ 1}{ 2}\tv^2
   - \frac{ 1}{ 2}\tu_x^2
   + \frac{ 1}{ 2}\veps\cdot\left[
      \tu\tu_x^2 + 2\tu\tu_x\tv + \tu\tv^2 \right]
\Bigr)\,\Id x.
\]
The Miura transformation $\cE_\veps\to\cE_0$ is given through
\begin{equation}\label{MiuraKBeKB}
u=\tu+\veps\cdot\bigl(\tu_x+\tv\bigr),\quad
v=\tv+\veps\cdot\bigl(\tu\tu_x+\tu_{xx}+\tu\tv+\tv_x\bigr).
\end{equation}
The recurrence relations upon densities of the Hamiltonian functionals
for~\eqref{KB} are
\begin{eqnarray}
\tu_0&=&u,\qquad
\tv_0=v,\qquad
\tu_k=-D_x(\tu_{k-1})-\tv_{k-1},\nonumber \\
\tv_k&=&-D_x^2(\tu_{k-1}) - D_x(\tv_{k-1})
   - \sum_{\ell+m=k-1}
      \bigl[\tu_\ell D_x(\tu_m) + \tu_\ell \tv_m\bigr],\quad k>0.
      \label{KBHamRec}
\end{eqnarray}
Relations~\eqref{KBHamRec} do not produce any auxiliary trivial
conserved densities.
\end{theor}

Relations~\eqref{KBHamRec} provide the formulas for the Hamiltonians
of the (twice-{} and thrice-) modified Kaup\/--\/Boussinesq
equations~\cite{PavlovJNlin}. The modified fields are conserved;
nevertheless, we note that they are not obtained from the Hamiltonians
for~\eqref{KB} by the Miura substitutions and that the
modified fields cannot be used as the negative terms in the general
scheme of~\eqref{KBHamRec}.

Substitution~\eqref{MiuraKBeKB} provides the canonical factorization of
the trivially extended Poisson structure
$\hat{A}_{1}+2\veps\hat{A}_{2}$ for the Kaup\/--\/Boussinesq
equation, where~$\hat{A}_1$ is defined in~\eqref{FirstStr} and
\[  
\hat{A}_{2} = \left(\begin{array}{cc}
D_x & \frac{1}{2} D_x\circ u\mathstrut \\
\frac{1}{2}u\,D_x & D_x^3+\frac{1}{2}D_x\circ v+\frac{1}{2}vD_x\mathstrut
\end{array}\right).
\]  
The two sequences of Hamiltonians, see~\eqref{KBHamRec},
for the deformed Kaup\/-\/Boussinesq
equation~\eqref{KBe} are inherited from the original
functionals for the Kaup\/--\/Boussinesq equation~\eqref{KB}
by using the Miura substitution~\eqref{MiuraKBeKB}.
The correlation between the higher flows for~\eqref{KBe}
and the symmetries of~\eqref{KB} is standard,
see~\cite{KuperIrish} and references therein.

Taking the flow at~$\veps$ in extension~\eqref{KBe},
we obtain the adjoint Kaup\/--\/Boussinesq equation
\begin{eqnarray}\nonumber
u_\tau&=& uu_{xx} + u_x^2 + u_xv + uv_x,\\
v_\tau&=& - \bigl(2u_xu_{xx} +uu_{xxx} + u_xv_x + uv_{xx} - vv_x\bigr).
  \label{KBConj5}
\end{eqnarray}
System~\eqref{KBConj5} is bi\/-\/Hamiltonian w.r.t.\ two compatible
local Poisson structures~$\Gamma_1$ and~$\Gamma_2$,
where~$\Gamma_1$ is~\eqref{FirstStr} and
\begin{equation}\label{PoissonAdj}
\Gamma_2=\left(\begin{array}{cc}
0 & D_x\circ u_x + D_x\circ v\mathstrut \\
u_x D_x + v D_x &
  \begin{array}{l} -u_{xx}D_x - D_x\circ u_{xx} -{\mathstrut}\\
  {}\quad {}-v_xD_x-D_x\circ v_x\end{array}
\end{array}\right)
\end{equation}
We conclude that the adjoint Kaup\/--\/Boussinesq
equation~\eqref{KBConj5} is completely integrable.

\begin{rem}
Both Poisson structures~\eqref{FirstStr}
and~\eqref{PoissonAdj} for~\eqref{KBConj5} are
of differential order~$1$. This indicates that the system can be further
extended with a higher order symbol such that its complete integrability
is preserved. The situation is analogous to the Gardner extension
of the Korteweg\/--\/de Vries equation~\cite{Gardner}; we recall that
the extension of KdV resulted in the dispersionless modified KdV equation
whose Poisson structures are of order~$1$.
\end{rem}

\begin{rem}\label{SecKBMinus}
The `minus' Kaup\/-\/Boussinesq equation
$u_t=uu_x+v_x$, $v_t=(uv)_x-u_{xxx}$
admits a unique real quadratic extension
\begin{eqnarray}\nonumber
\tu_t&=&\tu\tu_x+\tv_x + \veps\cdot\bigl(\tu\tu_{xx}+\tu_x^2\bigr),\\
\tv_t&=&{(\tu\tv)}_x-\tu_{xxx}
    - \veps\cdot\bigl(\tu_x\tv_x+\tu\tv_{xx}\bigr)
    - \veps^2\cdot\bigl(\tu^2\tu_{xxx}+\tu\tu_x\tu_{xx} + \tu_x^3\bigr).
    \label{KBMe}
\end{eqnarray}
System~\eqref{KBMe} is assinged by the first structure~\eqref{FirstStr}
to the Hamiltonian
\[
\cH'(\veps)=\int\Bigl(
   \frac{ 1}{ 2}\tu^2\tv + \frac{ 1}{ 2}\tv^2
   + \frac{ 1}{ 2}\tu_x^2
   + \veps\cdot \tu\tu_x\tv
   + \frac{ 1}{ 2}\veps^2\cdot\tu^2\tu_x^2\Bigr)\,\Id x.
\]
The invertible Miura transformation from the deformed equation
to~\eqref{KB} is
\begin{equation}\label{KBMMiura}
\{u=\tu,\ v=\tv+\veps \tu \tu_x\}
 \qquad \Longleftrightarrow\qquad
\{\tu=u,\ \tv=v-\veps uu_x\}.
\end{equation}
The recurrence relation obtained from~\eqref{KBMMiura} provides only
the Casimirs $\int\!u\,\Id x$ and $\int\!v\,\Id x$; all the conserved
densities $\tv_k$ are trivial if $k>0$. Therefore
deformation~(\ref{KBMe}--\ref{KBMMiura})
of the `minus' Kaup\/--\/Boussinesq equation is trivial.
\end{rem}

\subsection*{Acknowledgements}
The authors thank E.\,V.\,Ferapontov, B.\,A.\,Kupershmidt, and
A.\,V.\,Mikhailov for helpful discussions
and are also grateful to M.\,Brodovsky for computer programming.
The content of this paper was reported at the $5$th Workshop
on Quantization, Dualities, and Integrable Systems at Pamukkale
University. The research was partially supported by
the Scientific and Technological Research Council of Turkey (TUBITAK).
A part of this research was carried out while A.\,V.\,K. was
visiting at Middle East Technical University (Ankara).


\begin{thebibliography}{99}\normalsize

\bibitem{Gardner}  
\by{Miura R M, Gardner C S, and Kruskal M D} 1968
   Korteweg\/--\/de Vries equation and generalizations.~II.
   Existence of conservation laws and constants of motion
\jour{J.~Math.\ Phys.} \vol{9} 1204--1209.

\bibitem{KuperAlgModel}
\by{Kupershmidt B A} 1982
   On algebraic models of dynamical systems
\jour{Lett.\ Math.\ Phys.} \vol{6} no.2, 85--89.

\bibitem{KuperIrish}
\by{Kupershmidt B A} 1983
   Deformations of integrable systems
\jour{Proc.\ Roy.\ Irish Acad.} \vol{A83} no.1, 45--74.

\bibitem{Fordy}
\by{Fordy A P} 1983
   Projective representations and deformations of integrable systems
\jour{Proc.\ Roy.\ Irish Acad.} \vol{A83} no.1, 75--93.

\bibitem{Mathieu}
\by{Mathieu P} 1988
   Supersymmetric extension of the Korteweg\/--\/de Vries equation
\jour{J.~Math.\ Phys.} \vol{29} no.11, 2499--2506.

\bibitem{Andrea}   
\by{Andrea S, Restuccia A, and Sotomayor A} 2005
   The Gardner category and non\/-\/local conservation laws for $N=1$
   super KdV
\jour{J.~Math.\ Phys.} \vol{46} no.10, 103517, 11~p.

\bibitem{MathieuN=2}
\by{Labelle P and Mathieu P} 1991
   A new $N=2$ supersymmetric Korteweg\/--\/de Vries equation
\jour{J.~Math.\ Phys.} \vol{32} no.4, 923--927.

\bibitem{Manin}
\by{Manin Yu I} 1978
   Algebraic aspects of non\/-\/linear differential equations
\jour{Itogi Nauki i Tekhniki.\ Ser.\ Sovremennye Prob.\ Mat.}
\vol{11} 5--151; transl.\ in \jour{J.~Sov.\ Math.} \vol{11}  
1--122.

\bibitem{PavlovFPM}
\by{Pavlov M V} 2004
   The Boussinesq equation and Miura\/-\/type transformations,
\jour{Fundam.\ Prikl.\ Mat.} \vol{10} no.1, 175--182.

\bibitem{PavlovJNlin}
\by{Pavlov M V} 2002
   Integrable systems and metrics of constant curvature,
\jour{J.~Nonlin.\ Math.\ Phys.} \vol{9} suppl.1, 173--191.

\bibitem{ClassSym}
\by{Bocharov A V, Chetverikov V N, Duzhin S V \textit{et al}} 1999
\book{Symmetries and Conservation Laws for Differential Equations of
Mathematical Physics} AMS Providence RI.
Krasil'shchik I and Vinogradov A (eds).

\bibitem{JKIgonin}
\by{Igonin S and Krasil'shchik I S} 2003
   On one\/-\/parametric families of B\"acklund transformations
\jour{Adv.\ Stud.\ Pure Math.} \vol{37} 99--114.

\bibitem{DeformLiou}
\by{Kiselev A V} 2002
   On B\"acklund autotransformation for the Liouville equation
\jour{Vestnik Moskovskogo Univ.} \vol{6} 22--26.

\bibitem{NonAbel}
\by{Kiselev A V and Golovko V A} 2004
   On non\/--\/abelian coverings over the Liouville equation
\jour{Acta Appl.\ Math.} \vol{83} no.1-2, 25--37.

\bibitem{Kiev2005}
\by{Kiselev A V and Wolf T} 2006
   Supersymmetric representations and integrable fermionic extensions
   of the Burgers and Boussinesq equations
\jour{Symmetry, Integrability and Geometry\textup{:} Methods and
Applications} (SIGMA) \vol{2} no.030, 19~p.

\end{thebibliography}
\end{document}